\documentclass[preprint]{IEEEtran}
\IEEEoverridecommandlockouts
\usepackage{cite}
\usepackage{amsmath,amssymb,amsfonts}
\usepackage{algorithmic}
\usepackage{graphicx}
\usepackage{textcomp}
\usepackage{xcolor}
\usepackage{array}
\usepackage{tabularx}
\usepackage{comment}
\usepackage{soul}
\usepackage{longtable}
\usepackage{lscape}
\usepackage{hyperref}
\def\BibTeX{{\rm B\kern-.05em{\sc i\kern-.025em b}\kern-.08em
    T\kern-.1667em\lower.7ex\hbox{E}\kern-.125emX}}
\begin{document}

\title{History-enhanced ICT For Sustainability education: Learning together with Business Computing students.\\
\thanks{The research was part-funded by the Royal Historical Society through the award of a Jinty Nelson Teaching Fellowship 2023-24 to the research team. }
}

\author{\IEEEauthorblockN{Ian Brooks}
\IEEEauthorblockA{\textit{School of Computing and Creative Technologies} \\
\textit{University of the West of England}\\
Bristol, UK \\
Ian.Brooks@uwe.ac.uk}
\and
\IEEEauthorblockN{Laura Harrison}
\IEEEauthorblockA{\textit{School of Arts} \\
\textit{University of the West of England}\\
Bristol, UK \\
Laura2.Harrison@uwe.ac.uk}
\and
\IEEEauthorblockN{Mark Reeves}
\IEEEauthorblockA{\textit{School of Arts} \\
\textit{University of the West of England}\\
Bristol, UK \\
Mark.Reeves@uwe.ac.uk}
\and
\IEEEauthorblockN{Martin Simpson}
\IEEEauthorblockA{\textit{School of Arts} \\
\textit{University of the West of England}\\
Bristol, UK \\
Martin.Simpson@uwe.ac.uk}
\and
\IEEEauthorblockN{Rose Wallis}
\IEEEauthorblockA{\textit{School of Arts} \\
\textit{University of the West of England}\\
Bristol, UK \\
Rose2.Wallis@uwe.ac.uk}
}

\maketitle

\begin{abstract}
This research explores the use of History to enhance education in the field of ICT For Sustainability (ICT4S) in response to a challenge from the ICT4S 2023 conference. No previous studies were found in ICT4S but the literature on History and Education for Sustainable Development is reviewed.  An ICT4S lecturer collaborated with History lecturers to add an historic parallel to each week’s teaching on a Sustainable Business and Computing unit for final year undergraduate BSc Business Computing students. A list of the topics and rationale is provided.  Student perceptions were surveyed before and after the teaching and semi-structured interviews carried out. A majority of students saw relevance to their degree and career. There was an increase in the proportion of students with interest in History. The paper explores the lessons learned from the interdisciplinary collaboration, including topic choice, format and perceived value. The project has enhanced the way we approach our subjects as computing and history educators. We believe this is the first empirical, survey-based study of the use of history to enhance ICT4S education. The team will extend the research to a larger unit covering a wider range of computing degrees. 
\end{abstract}


\section{Introduction}
What role can History play in the teaching of ICT For Sustainability (ICT4S)? The question was raised in the ICT4S EDU educators workshop in June 2023. A roundtable discussion surfaced the following questions and observations:   What would be the purpose of using History in ICT4S?  \textit{To see the errors of the past and not reproduce them. To question what is understood as progress. To learn from past experiences looking for similarities and wisdom. To develop a theory of change. To see how things could be different. To teach sufficiency.} How to engage Computer Science students with History? \textit{Using discussion and parallels. Establishing the value of historical understanding to their career. To recognise when they might be involved in developing exploitative systems (How would I know if I was building a slave ship?).  To use place-based examples e.g. Bristol Slavery / Tobacco / Armaments. To use technical examples e.g. history of automobiles and society, Industrial Revolution, Printing. To use databases e.g. Slave Voyages \cite{SlaveVoyagesConsortium2018}, Legacies of British Slavery \cite{CentrefortheStudyoftheLegaciesofBritishSlavery2024}.} What challenges might ICT4S educators face and how could these be addressed? \textit{Lack of expertise, knowledge and vocabulary. Working with Historians, Anthropologists, STS (Science and Technology) scholars.} Some specific sources and authors were mentioned \cite{Costanza-Chock2020,Frankopan2023,Bonneuil2016,Black2012}. 

One of the authors of this paper participated in that workshop and had observed on the basis of several years of teaching ICT4S \cite{Brooks2019} it appeared that many students could benefit from a greater awareness of how society has been different and how it has changed. In other words, a greater knowledge of history might help students appreciate the nature of change needed toward a more sustainable economy and society. This paper sets out initial findings of a pilot on the use of History in ICT4S teaching in a UK university, the University of the West of England.
One of the authors designed and teaches the unit Sustainable Business and Computing, a mandatory unit for 3rd year undergraduate students on the BSc Business Computing programme. The unit is an eleven-week taught course with three hours per week contact time, which has been running every Autumn since 2020. The unit learning objectives are as follows and an overview of the unit by week can be found in table 1.

\begin{enumerate}
    \item Demonstrate a detailed knowledge and understanding of the concepts and competing definitions of a sustainable business.
    \item Evaluate companies' sustainability using the Global Reporting Initiative framework.
    \item Analyse the negative impacts of ICT on sustainability.
    \item Discuss in detail case examples of the role of ICT in delivering sustainability benefits.
\end{enumerate}

The first six weeks introduce general concepts of sustainability, business impacts and reporting. The remaining weeks focus on the sustainability impacts of computing. The unit makes extensive use of the United Nations Sustainable Development Goals as a framing of sustainability and the Hilty \& Aebischer (2015) Lifecycle – Enabling – Systemic (LES) model for ICT4S impacts \cite{Hilty2015}.

Following the ICT4S EDU 2023 workshop, the unit leader collaborated with the co-authors who are all Historians in the same university. We were introduced through the University’s Knowledge Exchange for Sustainability Education (KESE) group, which has a representative from every Department. The aim was to explore in practice how History teaching could enhance the existing ICT4S unit.  Teaching ICT4S should help students to build a mental model of the impacts of technology and a theory of change. An understanding of history may help students develop a stock of examples of previous technology implementation and models of the process of societal change.  

\subsection{Positionality}

The authors are all white British or American academic lecturers / researchers. The team has a balanced proportion of gender representation. We recognise the risk of a developed country bias in our work and selection of historical cases. We welcome criticism from colleagues in countries experiencing the legacies of colonialism and the opportunity to collaborate on future iterations of this work. 

\section{Background and Related Work}

We reviewed the literature for sources which have relevance to the use of History in teaching ICT4S. Below, we start with an overview of relevant textbooks and literature from Science and Technology Studies, since it was specifically mentioned by a participant in the ICT4S EDU workshop, and from the History field. We then offer a review of the academic literature specifically covering both History and Education for Sustainable Development (ESD) based on keyword searching. Lastly we review the relevant professional guidance applying to Higher Education in the UK.  

The academic field of Science and Technology Studies (STS), or Science, Technology and Society, includes topics which are broadly relevant to the understanding of history and sustainability. Johnson and Wetmore (2021) is a useful primer on STS and includes leading writers in the field such as Latour \cite{Johnson2021}. The readings are likely to be useful in teaching STS specific units but the book does not write about teaching STS or its use in ESD.   Journals such as Science, Technology and Society are also a valuable source. However, it has only published one paper which mentions ESD, this is in the context of discussing UNESCO’s Encyclopedia Life Support Systems (UNESCO-EOLSS) \cite{Kotchetkov1998}. There are a number of papers which discuss the STS curriculum including \cite{Cutcliffe1990} which emphasises the connection with the History of Technology and the importance of historical context. There are a number of well-established STS programmes such as that offered by MIT \cite{MassachusettsInstituteofTechnology2024}. So STS sources provide us with useful context for History in ICT4S education but little which directly addresses the topic. 

History of Computing is a directly relevant sub-field and there are a range of useful textbooks such as \cite{ORegan2021}. More recent papers have explored some of the entanglements of computing with colonial slave economies, such as \cite{Whittaker2023} exploring the relationship of Babbage’s thinking with plantation operations.   The film Hidden Figures \cite{Melfi2017} brought a sustainability issue (gender and ethnic bias) of History of Computing to a wider audience. The History of Computing potentially provides us with content to include in ICT4S education but students may already have been taught History of Computing as part of their course. Also, the parallels for ICT4S may be relatively constrained compared to a broader field of History. 

The field of Environmental History is gaining increasing academic and public interest, a recent example being \cite{Frankopan2023}. Schama 2021 calls for a ‘deep environmental history’ which ‘marries natural and human history’ \cite[p~324]{Carr2022} and continues “of all our histories, that of the environment is perhaps the one where past and future most hopefully seed each other.” \cite[p~328]{Carr2022}.  

The journal Environmental History has been published by Chicago University Press since 1996, though its precursor Forest History Newsletter was started in 1954, and brings together scholars and practitioners from the humanities, sciences, and social sciences to ‘explore the changing relationships between humans and the environment over time’. \cite{UniversityofChicagoPress2024}  

Recent work, such as All We Want is the Earth: Land, Labour and Movements Beyond Environmentalism (2023) by Patrick Bresnihan and Naomi Millner, examines the history of modern environmentalism with a focus on key interventions made by feminist, anti-colonial, Indigenous, workers’ and agrarian movements \cite{Bresnihan2023}. At a specific local level, Brownlee (2011) provides a history of the environmental movement in Bristol \cite{Brownlee2011}. 

Environmental history is a key feature of many undergraduate history courses, and recent years has seen a proliferation of stand-alone MA programmes in Environmental Humanities and Environmental History (Bristol, Durham, Nottingham, Plymouth to name just a few).  

Environmental History inevitably deals with the introduction of technologies and their impact. Frankopan particularly mentions the development of technology to extract and exploit coal, starting an age of fossil fuels\cite{Frankopan2023}. “By 1850, some 18 million people in Britain used as much energy as 300 million in China” \cite[p~218]{Frankopan2023}. There is also a section on weather modification technology in the 1950s both for benefit for farmers and as a potential weapon. Frankopan draws attention to a 1955 article “Can We Survive Technology?” by Von Neumann on the topic, which stated “All experience shows that even smaller technological changes than those now in the cards profoundly transform political and social relationships” \cite[p~519]{Neumann1955}. The technology of the agricultural ‘green revolution’ is also questioned as it becomes clearer that it has “significant impacts both on ecosystems and on health” \cite[p~273]{Frankopan2023}.   However, specific coverage of ICT impacts is sparse. Frankopan does draw attention to the way in which the digital age has boosted consumption and the packaging and transportation of parcels \cite[p~290]{Frankopan2023}. 

Turning to a keyword-based literature review, a search for history and ICT4S on SCOPUS returns only one paper \cite{Seznec2022} which situates “digital music within a cultural history of industrialisation” but does not address questions of education. SCOPUS was chosen as it provides coverage of a wide range of disciplines and includes a large proportion of IEEE and ACM publications. 

So broadening the search, the meta-discipline of ESD should also be useful to our study. ESD addresses the teaching of sustainability across all disciplines but for this case we are focused on its application in computing and in history. A literature review using the search terms “history” AND {education for sustainable development} finds 88 sources on SCOPUS (27 Dec 2023).  8 of these are classified by SCOPUS as being in the Computer Science field. We imported the SCOPUS data into the Parsifal literature review tool and used this to coordinate a consistent team review of the literature.  The following is a summary of the themes we identified in this literature. 

\subsubsection{History IN ESD (16) or History OF ESD (40)} The majority of sources identified in this search were papers about the history of ESD and its development as a pedagogical approach. There were only 16 papers which related to our interest in the use of History IN ESD. This topic has attracted little research to date. The following themes were found within this subset of 16 papers. 

\subsubsection{Environmental legacies of history} An interesting paper on Invasion Biology discussed teaching around the topic of invasive species in South Africa. The paper recognises species “were introduced for a particular use and then established self-sustaining and expanding populations beyond the area of introduction, where they may have both positive and negative impacts”\cite{Davies2016} but doesn’t take the next step of giving the historical context for those doing the introduction and their ‘particular use’ \cite{Davies2016}. There are clearly teaching opportunities to use the history of invasive species as an illustration of the development of the modern world. 

\subsubsection{Disciplines which lend themselves to ESD teaching} Several papers research the academic fields which are particularly suited to teaching ESD. These include History and Geography \cite{Ortega-Sanchez2020}, Liberal education, Civics \cite{Sherren2008} and Indigenous ways of knowing \cite{Manitowabi2022}. There was also a somewhat inconclusive paper on the History discipline \cite{Hendriawan2019}.  These are disciplines which require a critical engagement with the processes of change and with primary sources. They are described as supporting the “social conscience of their students, and the promotion of the concept of active citizenship" \cite[p~7]{Ortega-Sanchez2020}. 

\subsubsection{Teaching Teachers} Four of the papers related to the training of teachers to deliver ESD. This included curriculum mapping \cite{Hendriawan2020}, interdisciplinary training \cite{Fernandez2023}, and integration into initial teacher training \cite{Diez2021}.  \cite{Ortega-Sanchez2020} noted that a significant obstacle is the integration of new content into an “overpopulated” curriculum.  

\subsubsection{Local history as a significant lever} A frequent theme is the power of rooting ESD in local history and local lived experiences \cite{Manitowabi2022}. Amongst other approaches, \cite{Chapman2024} describes the use of old postcards of the local area for students to research local historical change. \cite{Diez2021} uses family trees as the focus for historical change, particularly emphasising the female stories. \cite{Hensley2020} explores local stories and \cite{Grunberg2023} recounts ESD using history at World Heritage Sites.  \cite{Zen2022} described the particular effectiveness of Kominkans (community learning centre) in Okayama, as a non-formal setting for ESD and for connection to the history of local geographical features.  

\subsubsection{Learning from past failures} Two papers specifically mentioned the use of past failures as a spur to learning – from past local disasters \cite{Sevilla2023} and from sports corruption \cite{Constantin2022}.  

\subsubsection{Models} There was no common model across the papers. \cite{Zen2022} describes their Quintuple helix model (QHM), \cite{Fernandez2023} uses a ‘Big History’ model from \cite{Christian2018} to situate events in the long sweep of time. \cite{Diez2021} uses a feminist methodology with the example of researching female members in family tree. 

\subsubsection{Computer Science field} SCOPUS identified eight sources as from the Computer Science subject area. Only two of these related to use of History IN ESD. \cite{Constantin2022} used sports websites as a source of histories of corruption in sports as a foundation for ESD in anti-corruption. \cite{Ortega-Sanchez2020} appears to be included because their study of teachers included 8.6\% in a Technology subject but there is no specific analysis relating to Computer Science.

\subsection{Professional Guidance} 

As staff in UK Higher Education, we follow the professional guidance of the Quality Assurance Agency for Higher Education (QAA). The relevant subject benchmarks in this case are History (2022) \cite{QAAHistory2022}, Computing (2022) \cite{QAAComputing2022} and Education for Sustainable Development (2021) \cite{QAAESD2021}.  Discussion of the relevant guidance follows. 

\subsubsection{QAA Subject Benchmark: History 2022} \cite{QAAHistory2022}

The History subject benchmark contains no explicit discussion of the history of technology or computing. It does discuss the use of ICT for delivering online history teaching and for assessment e.g. creation of digital history projects.  The benchmark identifies the need for History students to be proficient in “retrieval, selection, interpretation, analysis and synthesis of information from extensive datasets, both analogue and digital” \cite[p~7]{QAAHistory2022}.

The benchmark includes three paragraphs on the topic of Sustainability and the contribution History can make to sustainability and achieving the UN Sustainable Development Goals (SDGs). There is specific mention of the contribution of Environmental History but also acknowledgement of public history, cultural history and economic history. The benchmark includes the following assertion about the subject: 

“history prepares students to meet sustainability needs and challenges through its inherent attention to issues of change, continuity and causation; its demand for multi-perspectivity and multi-factorial understanding of events, issues and problems; its appreciation of the relationships between economic, political, cultural, social and environmental factors and systems; its ability to observe processes taking place in more than one place and more than one time; and its acknowledgement that there are rarely simple answers to problems.” \cite[p~6]{QAAHistory2022}. 

These are clearly attributes which would be beneficial for computing students learning about ICT4S. 

It is worth noting that the Royal Historical Society, the Institute of Historical Research (IHR) and History UK do not currently have explicit policies on ESD / sustainability. The Historical Association, the UK national charity for history, which offers support, resources and events for teachers (primarily primary \& secondary) has done some work in this area, and supports ‘Teaching for Sustainable Futures’, an online professional development programme for teachers run by the Centre for Climate Change and Sustainability Education at University College London.  

\subsubsection{QAA Benchmark Statement on Computing (2022)} \cite{QAAComputing2022}  

The Computing benchmark has little explicit discussion of history beyond “ensuring that case studies used as exemplars or as assessments are drawn from a diverse range, highlighting global perspectives, featuring diverse cultures and communities, showcasing the historical and cultural integration of computing and the diverse people who do computing”.  There is a page in the benchmark devoted to sustainability and ESD. There is also a useful appendix with two and a half pages of examples of ESD approaches. The sustainability section states that Computing courses should lead to students developing “core personal and professional competencies to address the key societal challenges highlighted by the United Nations Sustainable Development Goals for 2030 in their future working lives” \cite[p~6]{QAAComputing2022}. Computing graduates will have an “appreciation of domain-relevant issues related to sustainable development and the emergent manner by which these issues might be addressed” \cite[p~7]{QAAComputing2022}. “Emergent manner” implies a surprisingly reactive approach to sustainability when an attention to history might enable more robust anticipation of issues. Besides this, there are also tools available to foster awareness of sustainability issues e.g. SusAF \cite{Betz2022}.  Outside of the page and appendix on sustainability the subject benchmark has relatively little mention of the topic, indeed amongst the six course outcome categories detailed, sustainability is only mentioned in one – Professional Practice - which lists “equality, diversity and inclusion (EDI) and sustainability” amongst a range of other factors. 

\subsubsection{QAA Education for Sustainable Development Guidance 2021} \cite{QAAESD2021}

This guidance states that ESD student skills include “Use historical knowledge and an understanding of the consequences of past actions to envision how futures may be shaped” \cite[p~25]{QAAESD2021}. The ESD guidance applies as a cross-cutting issue across all subjects, so it follows that we should be teaching Computing students so that they acquire this history-oriented skill despite it not being explicitly mentioned in the Computing subject benchmark. 

\subsubsection{The Office for Standards in Education, Children's Services and Skills (Ofsted)}, which is responsible for the inspection of school standards in England, draws attention to the significance of students developing ‘knowledge and cultural capital’ that “that pupils need to be educated citizens”  \cite[p~250]{Ofsted2024}.  Teaching History can be an important component of acquiring that cultural capital as discussed in \cite{Baker2013}.

\subsection{Research Questions}

In the light of the ICT4S 2023 challenge and the sparse literature, we set out the following research questions for our interdisciplinary collaboration to make a contribution to knowledge concerning use of History in ICT4S education: 

RQ1: What historical parallels can be used to enhance the teaching of ICT4S? 

RQ2: What relevance do students see in the addition of History to ICT4S teaching?

\section{Methods}

In this section we set out the methods used to undertake the research which included the recording of online videos, identification of resources for students and in-class discussion. The impact was evaluated with pre- and post- surveys, and semi-structured interviews with students. The research plan, participant information sheets, consent form, survey instrument and semi-structured interview plan were all submitted to the university’s ethics committee and approved before the research started. Explicit consent was obtained from all participants.  The surveys, semi-structured interview questions and participant information sheet are available on Zenodo  \url{https://zenodo.org/doi/10.5281/zenodo.11073804}. The following sections set out the methods used to address each of the research questions.

\subsection{RQ1: What historical parallels can be used to enhance the teaching of ICT4S?} 

The ICT4S lecturer provided a summary of the topics taught each week on the existing Sustainable Business and Computing unit. The Historians proposed historical parallel topics for each week. These were drawn from the Historians' existing areas of specialism in order to reduce the potential burden of developing new content. After one iteration the list was agreed. The final topics are listed in Table 1 along with a brief discussion of the reason for selecting each topic. The first week was an in-person talk from the team of Historians introducing themselves and the relevance of History to sustainability. For weeks 2 – 10 individual Historians recorded a 10-15 minute video introducing students to the specific topic and its relevance to sustainability. The videos were made available to students via Panopto, the university’s internal video platform, a few days before the timetabled lecture. In addition, links were provided to additional readings and sources on the topic, which were made available via Blackboard the university’s virtual learning environment. These resources and links are included in the Zenodo package \cite{brooks_2024_11073805}. Access statistics were set to record the number of accesses made to the videos and materials. Each week an email announcement was sent to students with a link to that week’s video and materials.  Week 7 had an additional in-person question session with one of the Historians.  The content was not mandatory for students on the unit and was not part of the assessment as this was a research project. 

The effectiveness of the combined, videos plus in-class, intervention was tested in a survey of students at the start and end of the unit by an unprompted recall question about the historical topics students recalled. It was also tested in semi-structured interviews with a smaller sample of the students after the end of the unit, which included an unprompted topic recall question. The semi-structured interviews also included a question on the student’s view of the effectiveness of the format used to provide the content. 

\subsection{RQ2: What relevance do students see in the addition of History to ICT4S teaching?} 

Whilst the professional guidance literature is clear about the value and relevance of History in ESD, we also wanted to understand students’ perceptions of the relevance. The pre- and post- surveys included questions on perceived relevance of the History content to student’s degree and career, and their level of interest in history. The semi-structured interviews included questions about perceived relevance and about whether the content should be continued for future students on the unit.  

Surveys were administered online using Qualtrics and as paper copies in-class for students who did not complete online. A Qualtrics distribution was used to align pre- and post- responses so that respondent change in perception could be measured.  Semi-structured interviews were conducted online using MS Teams with automatically generated transcripts. Errors in the automated transcripts were corrected by the researchers. Transcripts were then analysed in the NVivo qualitative research tool to identify themes arising.  

\section{Results and Discussion}

\subsection{Results}

This section sets out the results of the questionnaire, semi-structured interviews and content access statistics. The questionnaire was administered in October 2023 at the start of the unit teaching and again in December 2023 – January 2024 after the end of the teaching. Semi-structured interviews were carried out in December 2023 – January 2024. 

\subsubsection{Questionnaire results}
21 students were registered on the unit. 17 students responded to the questionnaire at start of the unit and 11 respondents (65\% of the initial respondents) after the end of the teaching. Below are the results for each question. Since the cohort is small we have chosen to report the results in a qualitative fashion rather than evaluate statistical significance. We have not differentiated the results between students who watched videos and those who didn't since most students saw the Historians in class twice and some of the video views were anonymous and so the data sets do not match completely. 

\begin{itemize}

\item \textit{Q1 How relevant is a knowledge of history to your degree?}

There was a 12 pp increase in the percentage of students rating history as very relevant to their degree as shown in Figure~\ref{fig:Q1-1}. However, there was no overall change in the percentage for very or somewhat relevant to their degree.  There was a small shift from somewhat / very irrelevant before to neutral after (6 pp). At an individual level the results were mixed with three students expressing an increased relevance, two finding it less relevant and the majority (6) having an unchanged opinion as shown in Figure~\ref{fig:Q1-2}. 

\begin{figure}
    \centering
    \includegraphics[width=1\linewidth]{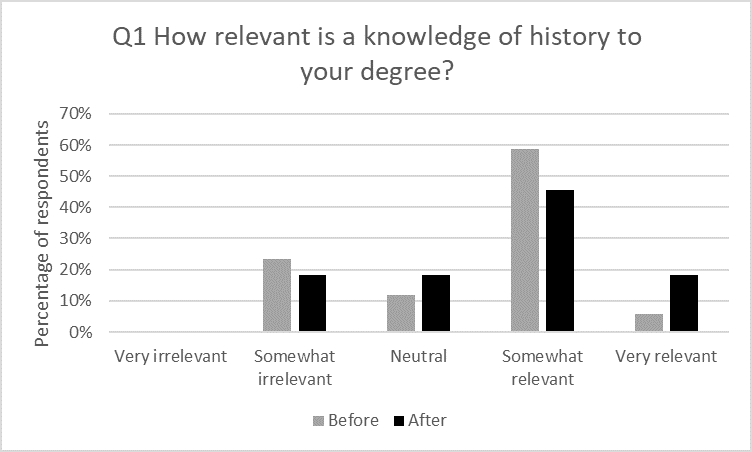}
    \caption{Relevance to Degree - All}
    \label{fig:Q1-1}
\end{figure}

\begin{figure}
    \centering
    \includegraphics[width=1\linewidth]{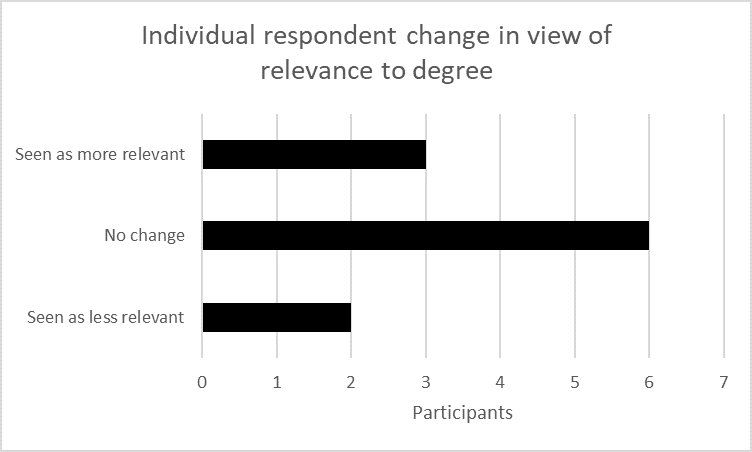}
    \caption{Relevance to Degree - Individual}
    \label{fig:Q1-2}
\end{figure}

\item \textit{Q2 How relevant is a knowledge of history to your career? }

There was a very small increase in the percentage of students rating history as very or somewhat relevant to their career (2 pp) as shown in Figure~\ref{fig:Q2-1}. In this case there was also a small shift from somewhat / very irrelevant before to neutral after (6 pp). At an individual level, two students came to see it as more relevant and three as less relevant as shown in Figure~\ref{fig:Q2-2}. 

\begin{figure}
    \centering
    \includegraphics[width=1\linewidth]{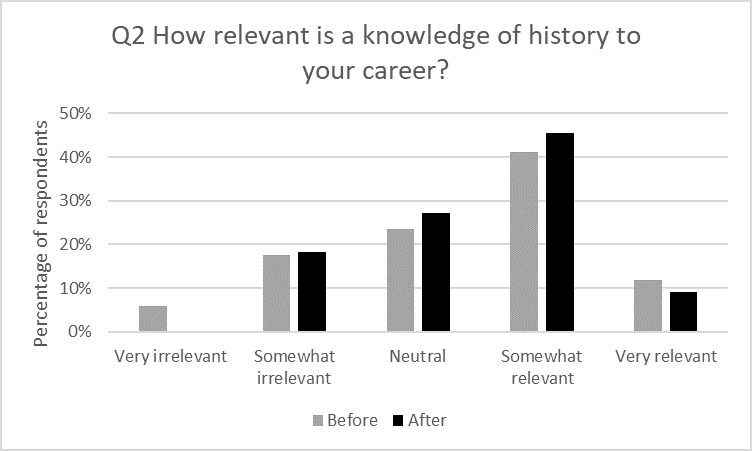}
    \caption{Relevance to Career - All}
    \label{fig:Q2-1}
\end{figure}

\begin{figure}
    \centering
    \includegraphics[width=1\linewidth]{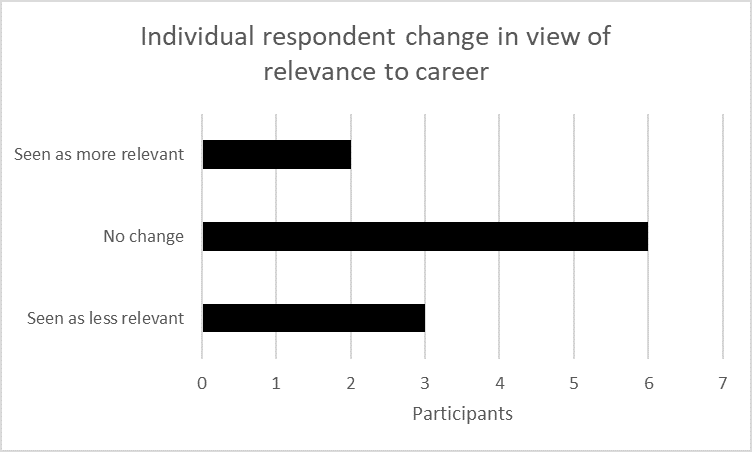}
    \caption{Relevance to Career - Individual}
    \label{fig:Q2-2}
\end{figure}

\item \textit{Q3 How interested are you in history? }

There was a noticeable increase in the percentage of students stating they are very or somewhat interested in history from 65\% before to 73\% afterwards as shown in Figure~\ref{fig:Q3-1}.  There was a stronger trend away from somewhat / very uninterested from 12\% before to none afterwards. At an individual level, most had an unchanged view (8) with 2 more interested and 1 with less interested than before as shown in Figure~\ref{fig:Q3-2}.  

\begin{figure}
    \centering
    \includegraphics[width=1\linewidth]{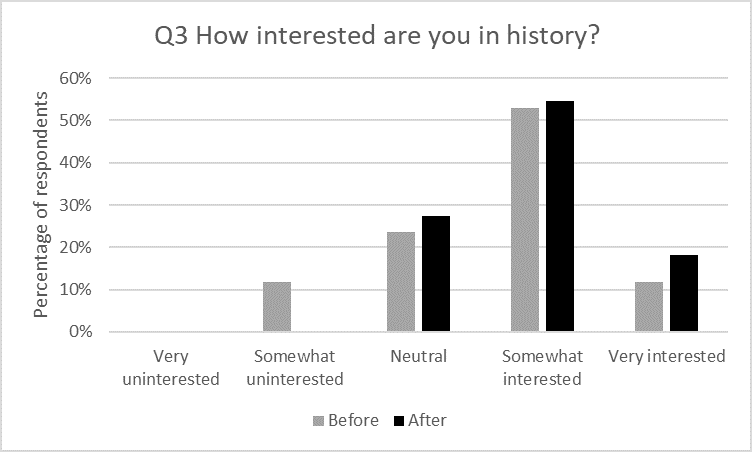}
    \caption{Interest in History - All}
    \label{fig:Q3-1}
\end{figure}

\begin{figure}
    \centering
    \includegraphics[width=1\linewidth]{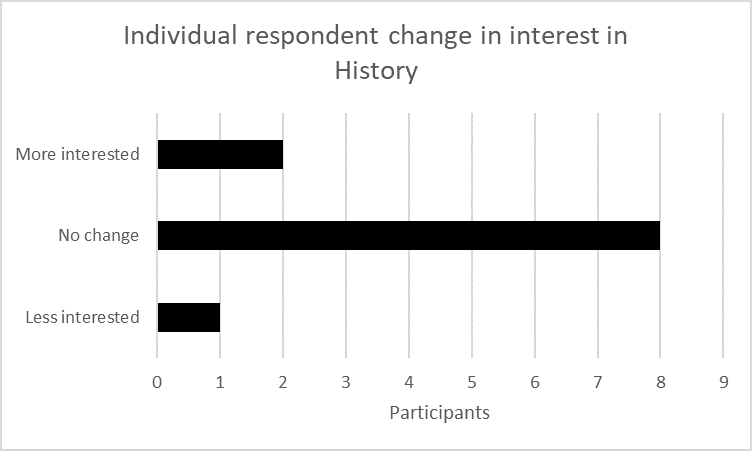}
    \caption{Interest in History - Individual}
    \label{fig:Q3-2}
\end{figure}

\item \textit{Q4. What historical events are you aware of which are relevant to computing and sustainability? }

In the pre-questionnaire, 5 out of the 7 who answered mentioned events relating to the history of computing and the internet. In the post-questionnaire the pattern of responses changed with 6 out of 7 also referring to broader historical events and the UN SDGs in addition to computing events. 
\end{itemize}

\subsubsection{Interview results}
Semi-structured interviews were carried out with 4 respondents to explore further the opinions of students. The four students volunteered in response to an appeal for interviewees, so the sample is self-selecting, may not be representative and includes students who generally had more engagement with the history project. The transcripts were analysed using NVivo and the themes are described below. 

\begin{itemize}

\item \textit{Topics} From the eight topics covered in the videos, these are the unprompted recollections of the participants: What makes a business dirty, the rise of Environmentalism (Participant 1).  Match Girls strike, Slavery abolition, East India Company, London used to be a very dirty place, Silent Spring (Participant 2).  Dirty Business, East India Company, Match Girls, Medical Technology (Participant 3). Medical History (Participant 4). Students were also asked about what other history they would have wanted to hear about. “What I'd have loved to hear, maybe a bit of history of computing as well.” (Participant 2). “a lot of World War Two stuff is interesting” (Participant 4). 

\item \textit{History taught on Degree programme} Students reported no formal history content in the first two years of the course. They did mention one lecturer who had discussed a little history of computing in one unit, though it was not formally on the curriculum, and they recalled “the history of computers and how like what the first computer was made and how it was created” (Participant 1).   “bits on the history of the web” (Participant 2). “very, very little, little to none” (Participant 4) 

\item \textit{Format and accessibility} When asked about the format there was a range of opinions. On the length of the videos: “if the videos were longer, it would be cool” (Participant 1), “I personally don't mind, but I've heard some students say I would have loved if all of them were shorter.” (Participant 2), “a longer video would definitely lose my attention” (Participant 4). On videos versus face-to-face: “I think it's always better to do stuff in person” (Participant 1), “, I prefer the face to face” (Participant 4), “a combination of both would be good” (Participant 2), “I would prefer videos ... so we can watch it back whenever we need to” (Participant 3).  One participant complained that the videos were difficult to access though the University’s systems, especially if using a phone “It is kind of hard to access, the materials cause you have to go through like logging in and then clicking blackboard and then you have to click another link and link and link.”  (Participant 1). 

\item \textit{Relevance to degree} In discussing the relevance of the History content to their degree programme, students noted: “the green IT one I think it was like the rise of environmentalism that obviously again that links like I guess you could say perfectly with the module that it was associated with you know sustainable business computing that I would say is like I mean to be fair I would say all of them were all the byte sized like history videos all had like pretty relevant  information that can be linked to that module specifically.” (Participant 1), “of course it's primarily relevant to this module we are doing 'cause that touches on employees’ welfare which is part of the SDGs. It's also part touches on environment.” (Participant 2). 

\item \textit{Relevance to career} Students expressed views that the content had particular relevance to their future career choices: “But another thing which I thought about it's because we are obviously working for companies whether we like it or not. So it's kind of know a history of corporations as a whole. I think it's very relevant to my career, at least.” (Participant 2). “I think I'm gonna link back to the East Indian company. I feel like it got me contemplating the motivations behind corporate actions, and it also helped me to brainstorm the kind of companies or the kind of corporations that I would wanna work for and the exemplary practices that they need to uphold.” (Participant 3). Participant 4 mentioned that the industry they want to join has particular ethical risks around technology.  

\item \textit{Should this be continued for future students?} “yeah, I think it should be 100\% like implemented not just this module, but like I feel like every module really should have some elements of linking back to history” (Participant 1). “I personally think so. But again. I think it's more relevant for Year 3 'cause we kind of are more equipped to value this” (Participant 2). “Yes, definitely.” (Participant 3). 

\item \textit{Should it be mandatory?}  “maybe mandating it could really help some students“ (Participant 2), “if it's relevant for the assignments, then yeah, but other than that, I don't think it should be mandatory” (Participant 4)   

\end{itemize}

It is hard to provide a more concise summary than this comment from Participant 3: “I believe incorporating the history content into the ICT education helps us students to gain a comprehensive understanding of the past and the skills to navigate the future and creating links between ICT and history helps form a sort of an interdisciplinary approach, which is to understand the impact of technology on historical events and societal changes vice versa.”

\subsubsection{Access to videos and resources }

The access statistics for the videos showed low levels of access varying between 1 and 5 unique student views. The access figures for each video are shown in Table 1. It is possible for there to be under-counting if students watched videos together on one device. 

\subsection{Discussion}

The following section discusses the results and findings. 

\subsubsection{Relevance of History} The pre- and post- questionnaires showed a small (12 pp) increase in the percentage of students finding history very relevant to their degree and small (6 pp) shift from irrelevance towards neutrality. The majority of students did not change their views which is unsurprising as most students did not watch the videos. Of those who changed their view, slightly more came to view it as more relevant to their degree. As one student observed in the semi-structured interview, the History content was chosen to be relevant to the specific unit.  With respect to relevance to career, there was a small (6 pp) move from perceived as irrelevant towards neutral and relevant. Again, the majority of students did not change their view. The interviewees had much more to say in favour of relevance to career, particularly about how they would consider the companies they might want to be employed by. This was catalysed by their learning about the activities of the East India Company. There was a small (12 pp) increase in the level of reported interest in history before and after the unit.  

\subsubsection{Topics} Interviewees varied in their unprompted recall about the topics taught but between them they remembered all but two of the topics – Slumming it, The Nuclear Age. No single topic stood out as having higher recall than others. So the variety of topics and periods was effective for the diverse interests of the students. The finding that these students had very little teaching on the history of computing suggests that this unit ought to include a topic on this in future runs. The literature reviewed underlines the power of local history but the selected topics were not particularly local to the university city. This is an area of potential improvement for future runs of the unit. The technical hurdles were clearly one reason for the low level of access to the videos and resources but a larger reason seems to be that the content was optional and students focus on what is mandatory. There is a bind - low uptake if voluntary but risks being seen as a “chore” (Participant 1) if mandatory. The team will reconsider the optional / mandatory choice after running a similar exercise with a larger student cohort on a different teaching unit.  

\subsubsection{Format} There was not a consistent pattern of responses in the interviews – some preferring video, some in-class face to face.  There was a frustration from one interviewee about the difficulty of accessing the videos and another only became aware of the videos towards the end of the unit. Access to the videos is clearly an issue which we need to improve. We may be able to do this by including a direct link to the video in the teaching slides which students do typically access. A student suggested that the videos should be uploaded to YouTube with unlisted URL but the team are concerned that this would add burdens of content production and clearance. 

\subsubsection{Reflections by academics} The team members provided reflection on their learning from the collaboration. 

The ICT4S academic reported learning about various approaches to historiography and a sense of the power of primary sources. He was also trained out of a tendency to ask “what would you have done in that situation?”. The experience provided a fund of examples of how the world can be different and how it can be changed. He was surprised that, in a class discussion, none of the students were aware of the term ‘Luddite’ and their history, which seems particularly relevant for students on a technology course. He reflected that thinking about the history of slums and of public health made him aware that his teaching was often about sustainability impacts of businesses and that there is gap between that and sustainability of societies. The collaboration expanded his understanding of the mechanisms and pace of change e.g. comparing the timescale it took for transatlantic trade in enslaved peoples to go from acceptable, legal and profitable to unacceptable, illegal and interdicted, with the timescale it will take for the fossil fuel industry to go through a similar trajectory \cite{BrooksIan2023TitO}. Whilst there is a growing literature on sustainability in computing education \cite{Heldal2024,Peters2024} this does not yet appear to have integrated the power of history into such curricula.

The History academics, reflecting on the experience, reported that the collaboration has helped them to think more explicitly about how their work addresses the SDGs – and how History and ESD can mutually enrich one another. They have thought a lot about how the goals of ESD can be supported by incorporating History into the curriculum – one key next step might be to explore the reverse. As a collaborative project this has supported them in advancing their pedagogical approaches – not only as an additional experience of communicating history to a broader audience, but in bringing together different approaches, viewpoints and experiences. They also commented on encountering an industry with different expectations e.g. finding that the ICT sector has very low levels of unionisation and that unions are not perceived as necessary. The Historians also perceived that the anticipated career paths for computing students were more focused than those of history students. There was more content in the unit which specifically career-focused. This changed the nature of discussions about careers and employment expectations. 

\subsection{Answering the Research Questions} In summary, with respect to RQ1: What historical parallels can be used to enhance the teaching of ICT4S? We used a range of historical parallels which illuminated topics of direct environmental and social impacts as well as the actions of individuals, companies and governments. The topics with the highest unprompted recall were Dirty Business / East India Company, Match Girls Strike, Medical Technology and the Rise of Environmentalism. Nevertheless, six out of the 8 topics had unprompted recall. The topics are all from the modern period when the rapid expansion of technology and limited companies provide clear parallels with which students can relate. We could have included parallels from the History of Computing. Reflecting on the literature, we should have made more use of the local historical context.  Concerning RQ2: What relevance do students see in the addition of History to ICT4S teaching? The pre- and post- questionnaires showed some perception of relevance to their degree and career. In the interviews, those students who engaged with the content expressed positive views about the relevance and discussed how it had led them to ask more questions about the companies they might work for. They recommended that the content should be included in future runs of the unit.  

The small cohort and low levels of access to the videos and resources mean that we cannot make strong claims for the findings in this study. However, those students who engaged with the content found it worthwhile and saw relevance to their degree and career. The effort to produce the videos was relatively low, as the Historians were drawing on their existing expertise, and these resources can be used in future runs of the unit. All members of the teaching team have found the collaboration has enhanced their teaching in ICT4S, ESD and History. On this basis we conclude that the project was a useful contribution to our knowledge and to the students who chose to engage.  It has been a useful pilot with small class size, the research team have approval to extend the project to another unit - Information, Networks and Society - with 110 students, which should provide results with more statistical significance. 

\subsection{Guidance for Implementation}
Although this was a pilot exercise with a small cohort, we can formulate some guidance for ourselves and others considering a similar use of history to enhance their ICT4S teaching.
\subsubsection{Seek Historians} Historians bring their training in their discipline and experience in their period / topic. An interdisciplinary discussion around the unit content should readily generate relevant historical parallels which address unit learning objectives.
\subsubsection{Focus on reuse} Stick to topics where colleagues already have expertise and resources, unless there is development investment available for new research.
\subsubsection{Short video, follow-up resources and in-class} A 10-15 minute video can give an overview of the historic topic. Ensure that students can access the videos as easily as possible. Make readings and resources available if a student wants to explore further. Adding some in-class sessions enables interaction and response to students questions and observations.
\subsubsection{Local history} Where possible, add local history content to help students identify with the potential for place-based change.
\subsubsection{History of Computing} If the student cohort have not already been taught about the history of computing then consider adding specific computing-related history.
\subsubsection{Embrace uncertainty and complexity} Computing education often emphasises regularisation, simplification and deterministic solutions. In contrast, the historical content may show uncertainty, complexity and multi-factorial change. Embracing these differences can be of great value for ESD in Computing disciplines.

\subsection{Threats to Validity}

This research was with a small class size (21) so is not generalisable. The low level of access to the videos and materials means that many students did not study the History content and only encountered it in the two in-class sessions. The very small number of interviewees (4) were self-selecting and cannot be claimed to be representative.  

\section{Conclusion and Next Steps}

We report on an interdisciplinary research pilot bringing history content into teaching on ICT4S, specifically an undergraduate unit “Sustainable Business and Computing”. We believe this is the first empirical, survey-based study of the use of history to enhance ICT4S education. 

Whilst UK subject benchmark statements emphasise the value of History as a subject in Education for Sustainable Development, a literature review shows no published papers on using history in ICT4S teaching and very few on History IN ESD. What little has been published can be summarised along the following themes: fewer papers about History IN ESD (16) than History OF ESD (40); environmental legacies of history; History as a discipline which lends itself to ESD teaching; a quarter of papers were about Teaching Teachers; local history as a significant lever; learning from past failures; some description of models used but no clear winner. These topics will inform the next stage of this research, paying particular attention to the inclusion of local, place-based history.

A pre and post survey of students on the unit showed that a majority of students saw relevance of History to their degree and career. There was an increase in the proportion of students with interest in History. This was a small cohort and the findings are not statistically significant and are not generalisable. Semi-structured interviews revealed that some of the participants found great value from the added historical content and advised that it should be continued for future runs of the unit. We also found that there were perceived hurdles to accessing the content which we need to address. The teaching / research team benefitted from the collaboration and sharing insights from history, sustainability and computing.  The project has created a set of videos and resources which will be of value to future runs of the Sustainable Business and Computing unit. 

Based on the student feedback and our lessons learned, we  believe this pilot was worthwhile considering the level of effort required to develop the content. The team will extend the research to a larger unit - Information, Networks and Society - in order to increase the sample size and range of Computer Science degrees represented.  Based on the findings of the literature review we encourage ICT4S educators to explore local historical parallels in their locations.  The literature review should be extended to include the keyword ‘heritage’ as relevant studies do not all use the keyword ‘history’. We encourage History organisations to address their own contribution to sustainability. 

Sustainability is a field where interdisciplinary working is essential and it has been a valuable experience for the team to work together, learning from each other’s discipline, expertise and methods. It has been a privilege to see History speak to present Business Computing students whose careers will shape future systems. 

\section{Acknowledgments}

We thank the students on the Sustainable Business and Computing unit for their participation in this research.

\newpage

\begin{landscape}
\begin{table}[htbp]
\centering
\caption{Teaching and History topics by week}
\label{table:1}
\begin{tabular}{|p{1cm}|p{7.5cm}|p{11.5cm}|p{1.5cm}|}
\hline
Week  & Sustainable Business and Computing Unit topic  & Historical topic and Reason To add  & Unique Student views of video  \\
\hline
1 (3 Oct 2023)  & What's the problem? Introduction to sustainability concepts, definitions and problems. Climate and Ecological Emergency. Reasons for hope. UN Sustainable Development Goals (SDGs).  & Introduction: History \& Sustainability  & Live, in-class introduction  \\
\hline
2  & What makes a good business? The pressures on businesses to be dirty - Sordid Sectors, Impatient Investors, Distracting Duties, Vicious Value Chains, Externalised Expenses, Careless Consumers, Dirty Darkness. Pulling the Levers of Change. Sustainability reporting. Global Reporting Initiative (GRI). Greenwash. Critical reading of sustainability reports.  & Dirty Business: Limited Companies and Corporations. Most students in this class (and most people) will work for corporations in their post-graduate careers. Therefore, this session drew students’ attention to the longer history of corporations, including the form’s entanglement with the history of colonialism in South Asia. This is not meant to simplistically disparage corporations in students’ minds, but to encourage them to think carefully about how their potential employers engage with the wider world and how the products students may create may be used and circulated in the global economy.     & 3  \\
\hline
3  & Environmental impacts - Climate Emergency. IPCC AR6. Ending burning of fossil fuels. Net Zero, Carbon Management. Scope 1,2 and 3 reporting. GRI305. Comparing carbon footprints. Climate Adaptation.  & The Atlantic Economy and Abolition. This session drew explicit parallels between past moral campaigns centered on international economic activity (against slavery) and current political activism around economic activities central to world trade (e.g., against inequality, for decarbonisation). By looking at concrete datasets such as the Trans-Atlantic Slave Trade database and the Legacies of British Slavery project, students saw the scale of the trans-Atlantic slave trade’s impact on the British economy, and how activist-generated solutions sought to create political and moral change in the face of that economic power.     & 3  \\
\hline
4  & Environmental impacts – Ecosystems. Earth system concepts. Safe Operating Space for humanity. Circular Economy. GRI 300 series.  & Dirty Cities: Pollution and the Urban Environment. The aim of this class was to draw a historical parallel by focusing on the impact of industry on Victorian cities (eg. The Great Stink) – but also looking at ideas of pollution in a broader sense, as a moral issue. Students were also introduced to the digitised Charles Booth poverty maps.  & 2  \\
\hline
5  & Social impacts. Health, Education, Discrimination. Sustainability motivations and values. Doughnut Economy. GRI 400 series.   & Slumming It: ‘Improving’ the Urban Environment. This topic was another relatively straightforward historical parallel – and discussed the language of ‘slums’ and the process of slum clearance from the 19\textsuperscript{th} century to the present day – highlighting issues of forced evictions and demolitions that have often been carried out in the name of “progress”. Also offered some thoughts about how we might approach the past – looking at the actions of those who enacted slum clearance vs the experience of those who lived and worked in “the slums”.  & 5  \\
\hline
6  & Economic impacts. Impacts of inequality. Fair wages / Modern Slavery, Taxation, Corruption. Progress towards SDGs. B Corps. GRI 200 series.   & Women, children and sweated labour. The London Match Girls’ Strike of 1888 was chosen as a direct parallel to modern sweated labour, reminding students that this is not a new mode of production. It was also selected to consider the agency and activism of workers, and particularly those from often marginalised groups – in this instance young women. This was useful as many of the students were unfamiliar with trades unions and considering work in industries where unionisation is low. The case study also raised questions about the cyclical intersection of poverty, consumption, and production.  & 1  \\
\hline
7  & Green IT – Hilty \& Aebischer LES model. 1\textsuperscript{st} order Climate Impacts of IT.  Clean / Dirty Cloud. Energy efficient computing. Sustainable IT market. Tech company Net Zero progress. Adaptation - Climate impacts on IT.  & Silent Spring: The Rise of Environmentalism. The historical content explored the beginnings of the environmental movement – with a focus on Carson’s \textit{Silent Spring }and the development of Earth Day.   & 4. Additional in-class Q\&A session.  \\
\hline
8  & Green IT – Other 1\textsuperscript{st} order / Lifecyle Impacts. Complex supply chains. Mineral sources. WEEE. Circular economy for Electronics. Worker exploitation.   & Make do and Mend. This topic was intended to explore the ‘make do and mend’ campaign during the Second World War.  & No video due to illness  \\
\hline
9  & ICT4S, \#Tech4Good, GeSI (Global Enabling Sustainability Initiative). 2\textsuperscript{nd} \& 3\textsuperscript{rd} order impacts. IT and decarbonisation. IT for UN SDGs. \#Tech4Good market.   & The Medical Revolution. This topic was intended to explore the development of medical knowledge and the practice of medicine in the eighteenth and nineteenth centuries. Different ways of thinking about the history of medicine and asking questions about a linear history of medicine as the development of \#Tech4Good were raised.    & 2  \\
\hline
10  & \#Tech4Bad. Dieselgate, Autonomous Weapons, Gig economy models, Social Media. The Tech that enables dirty businesses e.g. Fossil Fuels. AI X-risk versus current AI harms. Technology Assessment. Responsible Innovation. SusAF. Regulation and remedies.  & The Nuclear Age. This topic was intended to invite thinking about the ‘nuclear age’ inaugurated by the bombing of Hiroshima and Nagasaki. Questions were raised about policy, culture and campaigning, including recent debates about nuclear power as a green non-carbon solution. In this case the \#Tech4Good / \#Tech4Bad issue is not clear-cut.   & 2  \\
\hline
11  & Revision and career opportunities. Working in the sector - employers and jobs. LinkedIn networks.  & No video  &   \\
\hline
\end{tabular}

\end{table}
\end{landscape}

\newpage
\bibliographystyle{IEEEtran}
\bibliography{HistAndICT4SEDU}

\end{document}